\documentclass[twocolumn,caps,showpacs]{revtex4}

\usepackage{graphicx,color}
\usepackage{latexsym}
\usepackage{amsmath,amssymb}
\usepackage[draft=false]{hyperref}
\usepackage{mathrsfs}
\usepackage{comment}
\usepackage{mathrsfs}
\usepackage{multirow}
\usepackage{wasysym}

\DeclareSymbolFontAlphabet{\mathrsfs}{rsfs}
\DeclareMathAlphabet{\mathcal}{OMS}{cmsy}{m}{n}

\begin{document}

\title{Recognition of an obstacle in a flow using artificial neural networks}

\author{Mauricio Carrillo, Ulices Que, Jos\'e A. Gonz\'alez, Carlos L\'opez}
\affiliation{Laboratorio de Inteligencia Artificial y Superc\'omputo. Instituto de F\'{\i}sica y Matem\'{a}ticas, Universidad
	Michoacana de San Nicol\'as de Hidalgo. Edificio C-3, Cd.
	Universitaria, 58040 Morelia, Michoac\'{a}n,
	M\'{e}xico.}

\date{\today}

\begin{abstract}
In this work a series of artificial neural networks (ANNs) have been developed with the capacity to estimate an obstacle's size and location obstructing the flow in a pipe. The ANNs learn the size and location of the obstacle by reading the profiles of the dynamic pressure $q$ or the $x$-component of the velocity $v_x$ of the fluid at certain distance from the obstacle. The data to train the ANN, was generated using numerical simulations with a 2D Lattice Boltzmann code. We analyzed various cases varying both the diameter and position of the obstacle on $y$-axis, obtaining good estimations using the $R^2$ coefficient for the cases of study. Although the ANN showed problems for the classification of the very small obstacles, the general results show a very good capacity of prediction.
\end{abstract}


\maketitle

\section{\label{sec:Int} Introduction}

Obstructions in tubes transporting fluids are a problem of high impact in modern society, as they are present in people's daily life as well as in urban pipe networks \cite{Ugarelli}, public health \cite{Kernan} or industrial engineering \cite{Davies}.

In one hand the accelerated urban and industry growth in modern cities, implies that the obstructions in pipe networks are a very common problem, requiring a prompt reaction to solve it. These blockages can be caused by chemical or physical residues, as well as by structural defects of different sources. Pipes are one of the most usual way of transporting fluids in energy \cite{Sloan}, chemical \cite{Fraige}, manufacturing \cite{Datta, Wang} and water industries \cite{DuanHF}, as well as in houses, buildings and sewages \cite{Ashley, Bailey, Rodriguez}. Clearly, in all these cases, the requirements of an obstructed fluid delivery are crucial.

On the other hand, in the healthcare sector, the attention of medical problems due to obstructions and/or blockages in the innumerable conduits that transport biological flows around the human body are very frequent, and in many cases can be fatal for the patient. For biological conduits, blockages can be caused by previous surgeries, foreign bodies, infections and deformations among many others. Thus, cases such as obstructions in the digestive system \cite{Deloose, Shi} or in the cardiovascular system \cite{Chiu, Fujimoto}, are related to medical emergencies which in many cases may involve the use of invasive clinical procedures whereby they require prompt attention.

All these scenarios motivate the scientific interest in detection and comprehension of shape and location of objects blocking or obstructing the flows \cite{DuanW}. In this work, we use artificial neural networks (ANNs), as a flow pattern categorization in order to recognize these obstructions for a fluid flow in a two dimensional conduit. In this context, researches such in \cite{Ma, Lile, Massari}, have proposed different methodologies to identify obstacles, leaks or defects inside industrial or urban pipes. In particular ANNs as machine learning methods (ML) have been applied in problems of fluid dynamics mainly in flow phase pattern identification, for example in \cite{Naser, Rosa}. They have also been used as a tool for faster computational fluid simulations and turbulence prediction \cite{Ling, Bright, Lauret} or for defects classification in tubular structures using images \cite{Duran}.  

With this motivation, the objective of this work is to recognize the shape and location of an obstacle obstructing a pipe, whose dimensions were chosen considering pipes and networks that transport fluids for use in industrial and urban systems. For this, we have trained ANNs using physical information from the flow as input data. To than end, we have developed a generic 2D lattice Boltzmann numerical code \cite{Mohamad, Chen}, to simulate the flow of a fluid around an obstacle, contrasting the numerical solution with the benchmark \cite{Schäfer}. Our problem takes into account different scenarios: changing the diameter and location of the obstacle, viscosity and initial flow velocity, obtaining relevant physical information such as velocity, vorticity or dynamic pressure of the fluid along the numerical domain. In particular, we have considered the $x$-component of the velocity field of the fluid ($v_x$) or the dynamic pressure ($q$) as the fundamental information to be analyzed by the ANN. We chose a ratio between the width of the pipe and the immersed obstacle from 1/80 to almost a value of 1, this is relevant on physical scenarios where the flow is disrupted by obstructions that can be range from small blockages up to complete  of the pipe.

The content of the article is as follows: in section \ref{sec:one} we present a detailed description of the problem of blockage in the simulations with the lattice Boltzmann method (LBM, in section \ref{sec:two} we explain the methodology followed in the cases of study, describing the results obtained in section \ref{sec:Results}. Finally in section \ref{sec:Conclusions} we present the conclusions and future work.

\section{\label{sec:one} Numerical Simulations}

The simulation of the two dimensional flow around an obstacle, was performed constructing a numerical code based on the lattice Boltzmann method (LBM) as in \cite{Nuestro}. The LBM is very popular because it is easy to implement and it has a high capacity to perform computational simulations in a wide variety of physical problems \cite{Pan, Aidun, Yu1}, mainly applied in computational fluid dynamics \cite{Bouzidi, Feng}. 

\begin{figure*}
	\includegraphics[width = 18cm]{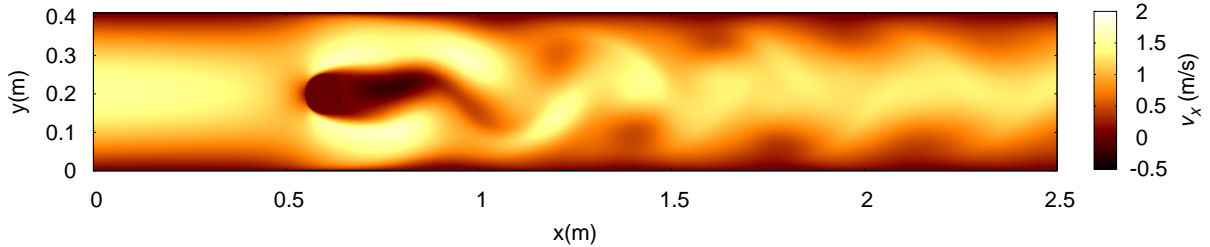}
	\caption{Magnitude of $v_x$ for the flow around a cylindrical obstacle with an income Poisseuille flow with a $v_c = 1.5$m/s. The obstacle is fixed at $x=0.6$m and the fluid flows over the x-positive direction along the pipe. Notice that the vortices formed after the obstacle are driven along the direction of the flow, generating the characteristic Karman vortex street.}
	\label{fig:Final_D21_vx}
\end{figure*}

For our cases of study, we consider a cylindrical obstacle immersed in an infinite medium (free flow), with a flow moving in the positive $x$ direction. The cylindrical obstacle is perpendicular to the $x-y$ plane, therefore is represented by a circle in the two dimensional simulation. We impose boundary conditions at the outlet of the flow far enough such that the characteristic flow parameters are not affected by the internal calculations \cite{Yu2}. For the solid walls of the pipe and obstacle, we employed a full bounce back boundary condition \cite{Mohamad}. The boundary condition representing the incoming fluid was set up with a given velocity profile as input for the numerical modeling, such that the flow after the obstacle presents a pattern related to the input velocity profile, the diameter of the obstacle and its location. 

We define $\beta$ as the value of the ratio of the diameter of the cylindrical obstacle and the width of the pipeline. The diameter of the obstacle is changed for values of $\beta$ ranging from $\beta = 0.0122$, representing small blocking elements, up to values close to $\beta = 1$, representing obstructions of almost the total size of the diameter of the pipe. For the domain of the simulation, a mesh of $165\times1,000$ nodes was used. Although the LBM code has dimensionless units, the system has been adapted to physical dimensions following reference \cite{Schäfer}. We have chosen the physical units such that the numerical domain corresponds to a total length of $L_y = 0.41$m in the vertical direction and $L_x = 2.5$m along the horizontal. We considered a Poiseuille income fluid flow in a stationary regime with a density of $\rho=10^{3}\rm{kg}/ \rm{m}^3$, an a kinematic viscosity of $\nu=10^{-3}\rm{m}^2/\rm{s}$, as used in \cite{Nuestro}. The location of all the studied obstacles  is at $x = 0.6$m of the pipe, and their position over the $y$-axis will be described in the following section.

We performed a wide amount of different numerical simulations, considering as free input parameters the inlet velocity profile, the diameter of the obstacle (changing the values of $\beta$), and the position of the obstacle with respect to the $y$-axis. An example of a numerical simulation is shown in Fig. \ref{fig:Final_D21_vx}, with an obstacle of size $\beta = 0.244$ and a Poiseuille flow with a characteristic velocity of $v_c=1.5$m/s. The numerical simulations were stopped until the system reached a neutral stability, which occurs before completing 30,000 iterations, or a physical time of approximately 16 seconds.

\section{\label{sec:two} Methodology}

In this work, we estimate the size and location of an obstacle measuring $v_x$ and $q$ after the obstacle. On this matter, one could propose the simplest case: considering a single sensor and applying a linear regression (LR) between the obstacle diameter versus the flow velocity at the location of the sensor. Although the LR is as good as the ANN in this case, the analyses on other scenarios show that the ANN outperforms the LR, and that is why we only present the ANN results. Moreover, the intention is to provide a first step to a methodology capable to estimate multiple and more complex obstacles or morphologies with non-symmetrical shapes. For this, we have defined a \textit{target region} around the area where the obstacle is immersed. On this region the size and position of the obstacle are estimated in terms of the proportion of the obstacle and fluid around, this will be explain in more detail further. We have also set different numerical sensors across the pipe on distinct measuring sites along the $x$-axis, an schematic representation of this is shown in Fig. \ref{fig:Schematics}. The measurement sites are located at $x = $1.75m, 2.10m, 2.45m, which we will refer as  A, B and C respectively.

With this in mind, the ANNs were selected because they are flexible in terms of input data and they are of easy implementation, outperforming linear regressions. In possible future works, we would like to increase the complexity of the problem, trying other, more sophisticated machine learning methods.

In next subsections we describe the cases of study, the structure of the ANN, the methodology used on the input and target data, as well as the selection of the training and validation sets, which are used to adjust the ANNs parameters, and the prediction set for which the results are shown.

\subsection{Database constructions and cases of study}

The ability of the ANNs, to predict the size and location of the obstacles, is tested in three different major cases:

\begin{figure*}
	\includegraphics[width = 18cm]{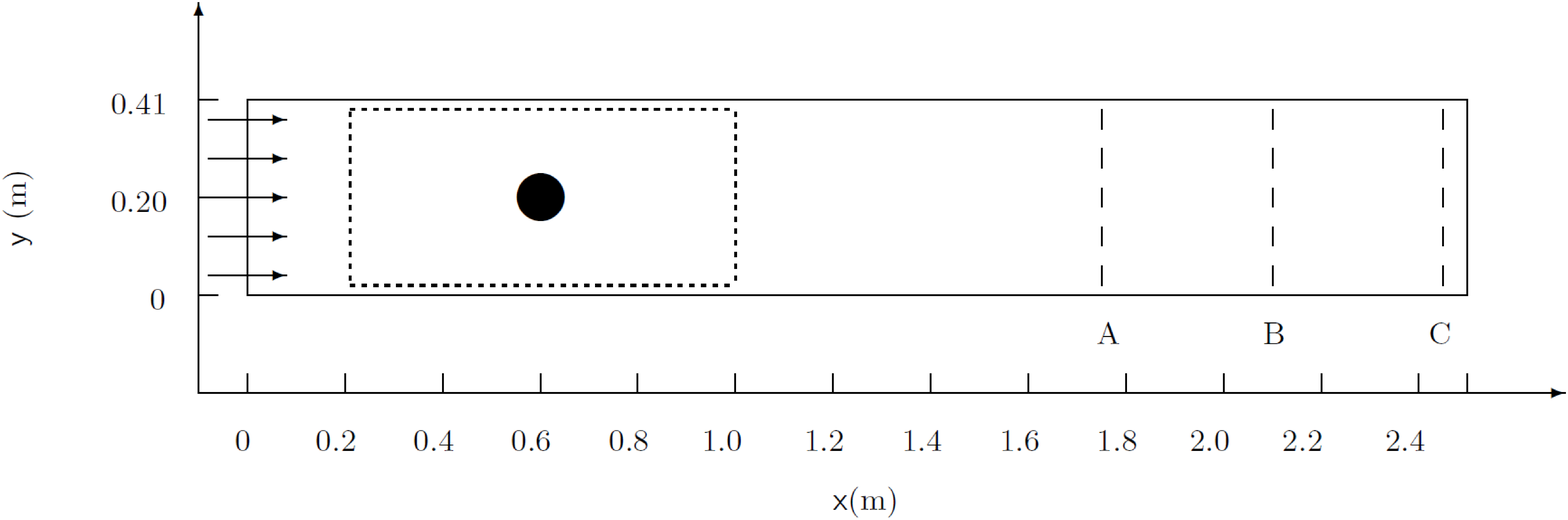}
	\vspace*{-1.25cm}
	\caption{A schematic representation of the 2D pipe, the cylindrical obstacle and different measuring sites across the pipe located at $\rm{A}=1.75\rm{m}, \rm{B}=2.10\rm{m}$, and $\rm{C}=2.45\rm{m}$, marked with dashed lines. The dot rectangle region illustrates the area used as target for the ANN.} 
	\label{fig:Schematics}
	
\end{figure*}

\begin{table}
	\begin{tabular}{|c|c|c|c|c|c|}\hline
		\multicolumn{6}{|c|}{Parameters of the studied cases} \\ \hline\hline
		{Case}   & Obstacles & $y$ Positions & Flow velocities & Space & Time \\ \hline
		1a  & 80  & 1  & 1  & 83  & 1  \\ \hline
		1b  & 80  & 1  & 1  & 1  & 300  \\ \hline
		1c  & 80  & 1 & 1  & 3  & 300  \\ \hline
		2a  & 4  & 43 & 1  & 83  & 1  \\ \hline
		2b  & 4  & 43  & 1  & 1  & 300  \\ \hline
		2c  & 4  & 43  & 1  & 3  & 300  \\ \hline
		3  & 11  & 1  & 12  & 10  & 1  \\ \hline
	\end{tabular}
	\caption{Parameters used in the simulations for the different cases studied. The column called ``Obstacles" represents the number of different diameters of the obstacles. The column ``$y$ Positions", gives the number of different locations of the center of the obstacles along the $y$-axis. The column defined as ``Flow velocities.", refers to the number of different income flow velocities used in scenario \ref{p:CaseIII}. Meanwhile, the column ``Space" is related to the number of equidistant sensors implemented along a measurement site. Finally, the column ``Time", corresponds to the number of time steps extracted in the numerical evolution performed on each considered numerical sensor.}
	\label{tab:ParametersCases}
\end{table}

\begin{enumerate}
	\item We perform 80 different numerical simulations for a two dimensional cylindrical obstacle immersed in the flow, located on the center of the $y$-axis, i.e., at the half of the pipe diameter. We simulate obstacles with diameters ranging from $d=0.005$m to $d=0.395$m in steps of 0.005m, i.e., values of $\beta$ ranging from $\beta=0.012$ to $\beta=0.964$ in steps of $0.0122$, plus a tiny obstacle of 0.001m. From these cases, we select as the prediction set those obstacles with diameters ranging from $d=0.02$m to $d=0.395$m in intervals of $\Delta d = 0.025$m. Meanwhile, the validation set consists of the obstacles from $d=0.025$m to $d=0.375$m with $\Delta d = 0.025$m also. The rest of the 49 obstacles are used as the training set. This case is divided into three different sub-cases:
	\begin{enumerate}
		\item The profiles of $v_x$ or $q$, at $t=16$ seconds at the end of the numerical evolution, are considered as input vector. For simplicity in terms of the ANN structure, the information extracted in only 83 of the 165 nodes of the numerical mesh. This approach is examined to study if extracting the physical data, such as $v_x$ or $q$, available at a fixed time and distance is enough to give a proper estimation of the obstacle's size. \label{p:CaseIA}
		\item The time evolution of $v_x$ or $q$ over 300 time steps, from $t=0$s to $t=16$s measured on a single sensor at the center of the pipe, i.e., $y=0.210$m, is considered as the input for the ANN, using the symmetry of the conduit. With this, we inspect the limits of the predictions considering the smallest number of sensors as possible, with the advantage that it can take measurements over a fixed lapse of time. \label{p:CaseIB} 
		\item The same procedure as case \ref{p:CaseIB}, but in addition we add 2 other equidistant sensors, i.e., we have three sensor located at $y = 0.105$m, $0.210$m and $0.315$m. With this case, we study if increasing the number of sensors, compared with the previous one, increases the predictions accuracy.
		\label{p:CaseIC}
	\end{enumerate} \label{p:CaseI}.
	
\begin{table}[]
	\centering
	\begin{tabular}{|c|c|c|}
		\hline
		\multicolumn{3}{|c|}{Parameters for case \ref{p:CaseIII}} \\ \hline \hline
		\multicolumn{1}{|c|}{$\beta$} & \multicolumn{1}{l|}{Velocity (m/s)} & \multicolumn{1}{|c|}{\begin{tabular}[c]{@{}c@{}}Sensor \\ Location \end{tabular}} \\ \hline
		0.0122 & 0.15 & 11 (0.025m) \\
		0.0976  & 0.30 & 27 (0.065m) \\
		0.1952  & 0.45 & 43 (0.105m) \\
		0.2928  & 0.60 & 59 (0.145m) \\
		0.3904  & 0.75 & 75 (0.185m) \\
		0.4880  & 0.90 & 91 (0.225m) \\
		0.5856  & 1.05 & 107 (0.265m) \\
		0.6832  & 1.20 & 123 (0.305m) \\
		0.7808  & 1.35 & 147 (0.365m) \\
		0.8784  & 1.5  & 163 (0.405m) \\
		0.9638 & 1.65 & \\
		& 1.8 & \\ \hline
	\end{tabular}
		\caption{Database for diameters, income flow velocities and sensor locations used on case 3. The last column defines the location of the ten sensors over the $y$-axis on the LBM mesh, in parenthesis are shown their equivalent values on physical units. The values for $v_x$ and $q$ were extracted from these numerical sensors on both $x = 0$m and at $x=2.10$m on the pipe, the latter corresponds to measurement site B.}
		\label{tab:case3}
\end{table}
	
	\item In contrast with case \ref{p:CaseI}, we consider only three different obstacle's sizes with values of $\beta= 0.122$, $0.244$ and $0.488$, this is,  $d=0.05\rm{m},0.1$m and 0.2m. On every simulation, we changed the position of the obstacle among 43 different position over the $y$-axis, such that the obstacle could be either nearby the center of the pipe or near its walls, providing a total of 139 simulations. For each size of the obstacle we selected 22 simulations for the training and validation sets, and the leftover 21 simulations, which are equally spaced over the $y$-axis, were used for the prediction set. In order to prove the capacity of the ANNs for the estimation of a completely unknown obstruction size we include in the prediction set 21 simulations equally spaced over the $y$- axis for with an obstacle diameter of $\beta = 0.366$. This scenario, was also divided by the three different sub-cases following the same description as in 1a, 1b and 1c. \label{p:CaseII}
	
	\item In this case, we analyze a situation similar to case \ref{p:CaseI} where the obstacle is located again at $y=0.210$m and $x=0.6$m. With the difference that we have changed the income flow with a characteristic velocity from $v_c=1.5$m/s, used in cases \ref{p:CaseI} and \ref{p:CaseII}, to different values ranging from $v_c=0.15$m/s to 1.8m/s, in steps of $\Delta v_c=0.15$m/s. This analysis is performed to examine an extension of case \ref{p:CaseIA}, with for different input flow velocities and eleven different obstacle's sizes as shown in Table \ref{tab:case3}. In addition, we also explored the behavior on the network by adding more information about the income fluid and fewer sensors than in case \ref{p:CaseI}. For simplicity, we consider ten equidistant values along the $y$-axis of $v_x$ or $q$ before the obstacle, at $x=0$m according to the schematics on Fig. \ref{fig:Schematics}, and other ten equidistant values of $v_x$ or $q$ on measurement site B, analogous as done before on case \ref{p:CaseIA}. With these arguments, the input data for the ANNs consists on 20 values for each one of the 132 simulations produced. In order to explore the performance of the network and its dependence on the choice of the samples sets for training, validation and prediction, we have now selected the patterns randomly. The validation and prediction sets have 20 patterns each one, meanwhile the training set has 92.
	\label{p:CaseIII}	
		
\end{enumerate}

All the cases of study simulated are summarized on Tables \ref{tab:ParametersCases} and \ref{tab:case3}.

\subsection{Target Region}

\begin{figure}
	\hspace*{-2.0cm}
	\includegraphics[width = 11cm]{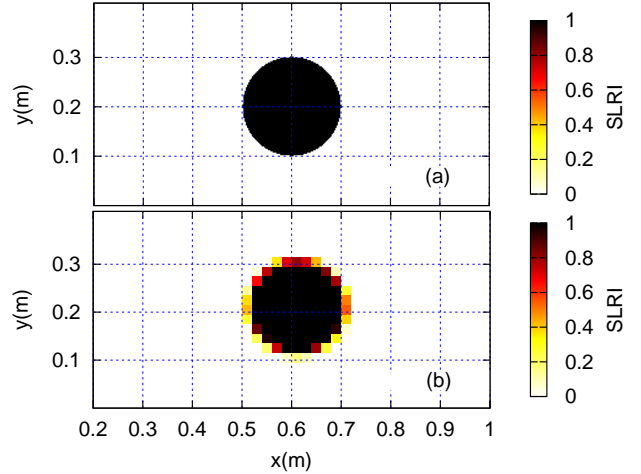}
	\vspace*{-1.3cm}
	\caption{In (a) plot is represented the solid/fluid elements on the LBM mesh simulation, while in (b) is the transformation from the simulation to the target grid. The lower number of cells used in the target grid in comparison with the LBM mesh causes a lost in resolution on the grid. The color box represents the solid/liquid ratio index (SLRI), where a 0 value means a cell is composed of only fluid elements and 1 represents a cell fully occupied by solid elements. The grid lines showed in the figures are only for reference and do not represent the size of a target cell.}
	\label{fig:TargetGrid}
\end{figure}

As we use ANNs trained with a supervised learning, we need to provide the targets, related to the true obstacle's size, shape and location. For this, we define a region inside the pipe containing the obstacle and its immediate surroundings as the target region. However, if we consider each one of the nodes of the LBM simulation on the target region to be the objective for the ANN, as illustrated by Fig. \ref{fig:Schematics}, it will have a huge number of outputs, making it computational expensive. Therefore, as a first approach we propose to consider a target region consisting in a fixed mesh of 40x20 cells, where each cell consists of 64 nodes of the numerical mesh. From now on we refer to the target region as the \textit{target grid}. 

In addition, we assign a numerical value for each cell, depending on the relation of nodes that represent fluid or solid elements on the numerical mesh. This means that the proportion of target cells occupied by the obstacle, is represented by the number of solid nodes divided by the total number of nodes of the numerical mesh contained in that cell. In other words, we have defined an occupation index for each cell of the target grid, where the relation solid/liquid of the cell was calculated, meaning that a value of 1 represents a cell containing only solid elements of obstacle and a value of 0 implies that there is only fluid in the cell. From now on, we will call this index the solid/liquid ratio index (SLRI). An example of the transformation from the LBM simulation of the numerical mesh to the target grid for an obstacle of size 0.2m is presented in Fig. \ref{fig:TargetGrid}, where the dark tones refers to the cells mostly occupied by the obstacle (high SLRI), whereas the cells with clearer tones mean a greater proportion of fluid elements (low SLRI). 

\subsection{Neural Network Structure} \label{sec:ANN_design}

Let us recall that for case $1a$, only half of the spatial nodes on the $y$ direction of the lattice are selected as inputs for the ANN, reducing the number of points where the vector fields are measured, from 165 to 83, simplifying the input data and the structure of the ANN and speeding up the computations. Therefore the input vector consists on 83 neurons, related with the profile of the fluid on the corresponding measurement site. For example, if we consider the $v_x$ values at the sensors on any measurement site, the input pattern is form by:
\begin{equation}
I = \lbrace v_{x_1}, v_{x_3}, \dots,v_{x_i},\dots, v_{x_{165}} \rbrace. \label{eq:83pts}
\end{equation}
where $i = 1,3,..., 165$ indexes the down-sampling made from 165 to 83 for the nodes on the LBM simulation mesh. For case \ref{p:CaseIB}, the input vector is defined by the time series of $v_x$ or $q$ at $y=0.21$m during 300 time steps represented by index ($t$), for example for $v_x$ this is:
\begin{equation}
I = \lbrace v_{x_{85},t_{1}}, v_{x_{85},t_{2}}, \dots , v_{x_{85},t_{300}} \rbrace. \label{eq:1pt}
\end{equation}
where $v_{x_{85},t}$ labels the value of $v_x$ at the node 85 of the simulation where $y=0.210$m, i.e., at the middle of the pipe at a certain time $t$. For case $1c$, the inputs of the ANN consist on the values of 300 time steps for $v_x$ or $q$ at the positions: $y = 0.105$m, 0.210m and 0.315m, on the considered measurement site. In Fig. \ref{fig:TimeSeries_3pts_vx} we present the time series evolution for $v_x$ at measurement site B. The inputs corresponding to this scenario are:

\begin{figure}
	\includegraphics[width = 9cm]{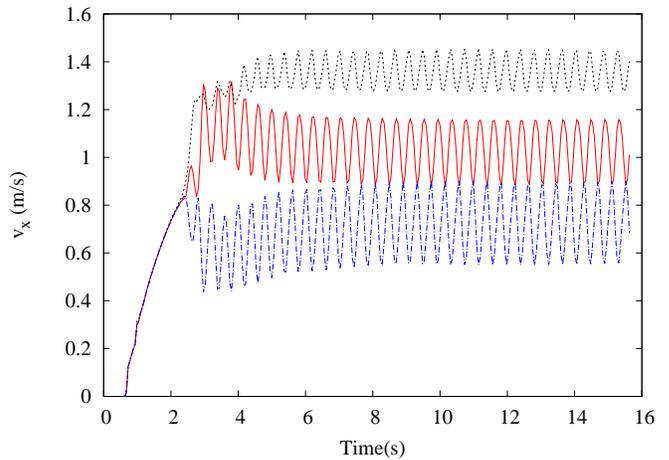}
	\caption{Time series of the $x$-component of the flow velocity at $y = 0.105$m (continuous), 0.210m (dashed) and 0.315m (dash-dotted) at measurement site B, for an obstacle of $\beta = 0.488$ centered at $y = 0.21$m.}
	\label{fig:TimeSeries_3pts_vx}
\end{figure}

\begin{multline}
I = \lbrace v_{x_{43},t_{1}}, v_{x_{85},t_{1}}, v_{x_{127},t_{1}},..., v_{x_{43},t_{300}},\\ 
 v_{x_{85},t_{300}}, v_{x_{127},t_{300}} \rbrace. \label{eq:3pts}
\end{multline}

On the second scenario, the inputs are the same as described in Eqs. (\ref{eq:83pts}-\ref{eq:3pts}). For case \ref{p:CaseIII}, let us recall that ten values on both at $x=0$m and $x=2.1$m (measurement site B) are considered as input pattern. For example, following table \ref{tab:case3}, an input pattern using $v_x$ is define as:
\begin{multline}
I = \lbrace v_{x_{11}}(x=0\text{m}),v_{x_{27}}(x=0\text{m}),..., v_{x_{163}}(x=0\text{m})\\
,..., v_{x_{11}}(x=2.1\text{m}),..., v_{x_{163}}(x=2.1\text{m}) \rbrace. \label{eq:case3}
\end{multline}

The same set of Eqs. (\ref{eq:83pts}-\ref{eq:case3}) are applied when the dynamic pressure $q$ is used instead of $v_x$ as input data for the network. \\
The internal structure of the ANN could differ internally on each case, i.e., the number of inputs and hidden neurons, but the number of outputs are constant for all the cases, since the objective is the same: approximate the shape of an obstacle blocking the flow. Recall that the region where the obstacle is located is described by 40x20 cells, which means that the target and the prediction has 800 elements. In a vectorized form, for an input pattern $I$, the result computed by the ANN is $\vec{O}=[O_1,O_2, \ldots ,O_{800}]$, and the $k$-th element of this vector is calculated by:
\begin{equation}
O_k=F_2\bigg(\sum^{J}_{j=1}\sigma_{jk} F_1\bigg(\sum^{I}_{i=1}\tilde{\sigma}_{ij} \tilde{I}_i+\tilde{\sigma}_{0j}\bigg)+\sigma_{0k}\bigg)
\end{equation}
where $1\le k\le800$ refers to the cell of the target grid; $\tilde{I}_i$ is the $i$-th element of the input vector and $J$ is the number of hidden neurons. $F_1$ and $F_2$ are the activation functions for the hidden and output layers respectively; $\tilde{\sigma}_{jk}$ and $\tilde{\sigma}_{0j}$ are the weights and bias terms between the input and hidden layer; $\sigma_{jk}$ and  $\sigma_{0k}$ are the weights and bias terms between the hidden and output layers.

The numerical implementation of the ANNs was developed from scratch using Fortran 90. Instead of using open source codes, we decided to use our own implementation to have full control of the details in the code, searching different structures and parameters in the learning process.
In one hand, the selection of the ANNs structure was done considering that it should be kept as simple as possible, in order to maintain computational advantage, and complex enough for its adaptation for unknown patterns. In our cases we found that ANNs with an input layer as defined in Eqs. (\ref{eq:83pts}-\ref{eq:case3}); one hidden layer with 20 neurons and an output layer with 800 neurons was complex enough to give good results without loss of performance for all the cases of study. All the ANNs used have hidden and output layers with sigmoid activation functions. On the other hand, the ANNs were trained using a \textit{backpropagation} algorithm \cite{Rumelhart}. In this work, we use this method to minimize a mean square error function, using a learning rate of 0.001, with a maximum of 15,000 iterations on training and using a cross-validation technique as stopping criteria. For clarity, all the results shown in here correspond to the prediction set. For more details about supervised training and backpropagation algorithm the reader can consult \cite{Bishop, Rojas}.

\section{\label{sec:Results} Results}

In order to estimate the ANN prediction accuracy for each one of the cases analyzed on the test set, we employ the $R^2$ coefficient. The calculation of the $R^2$ was performed over the target and predicted grid, considering the real and predicted SLRI. The $R^2$ is defined as follows:
\begin{equation}
	R^2 =  1 - \frac{\sum_{i=1} (O_i- \langle T \rangle)^2 }{\sum_{i=1}^{800}(T_i- \langle T \rangle )^2},
\end{equation}
where $T_i$ and $O_i$ are the $i$-th target and the ANN output respectively, and $\langle T \rangle$ is the average of the SLRI for all the target vector:
\begin{equation}
\langle T \rangle = \frac{1}{800}\sum_{i=1}^{800} T_i.
\end{equation} 
This means that the $R^2$ coefficient range is $(-\infty, 1]$, where a value of $R^2=1$ implies a perfect match term by term, between the target and the ANN prediction, while $R^2 \rightarrow 0$ means that the prediction approaches to $\langle T \rangle$.

\begin{figure}
	\includegraphics[width = 9cm]{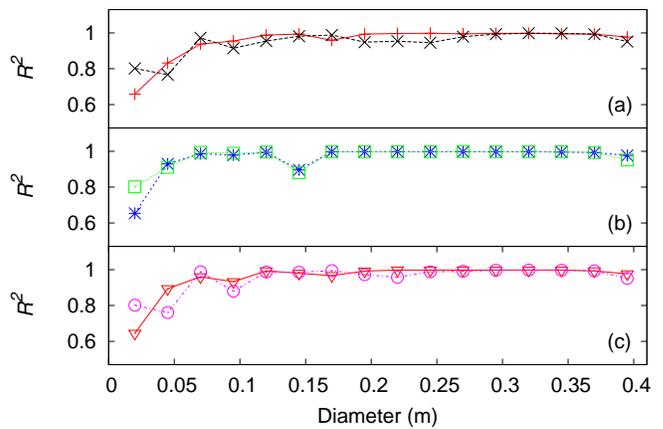}
	\caption{$R^2$ coefficient for predictions produced by the ANN for obstacles located at the center of the pipe with different diameters. The ANN is trained with data extracted with a sensor located over the line on B, and tested with measurements at A, B and C. On (a), the red curve (+) shows results using $v_x$ and $q$ is represented by the black curve ($\times$) on site A. In (b), the input data was extracted on measurement site B, and the blue ($\hexstar$) and  green ($\square$) curves represent the $R^2$ using the profiles $v_x$ and $q$ on respectively. In (c), the red ($\triangledown$) and magenta ($\bigcirc$) curves correspond to the results for $v_x$ and $q$ on measurement site C respectively. All $R^2$ coefficients are above 0.6 which can be interpreted as a good correlation between the prediction and target. Notice how the results are independent on the location of measurement.}
	\label{fig:Diams_Center_D_Test_DE_Velx_P}
\end{figure}

Let us recall that, the measurements of the fluid flow consist on one snapshot of the profile of $v_x$ or $q$ of the fluid, at the time when the system reaches a neutral stability. For the case described in \ref{p:CaseIA}, we present the results by measuring on detectors located not only on the training site B, but also on A and C as shown in Fig. \ref{fig:Diams_Center_D_Test_DE_Velx_P}. In this case, we study how the ANN behaves if it is only trained with information on measurement site B and tested on A, B and C, for the same obstacles.

\begin{figure}
	\hspace*{-2cm}
	\includegraphics[width = 11cm]{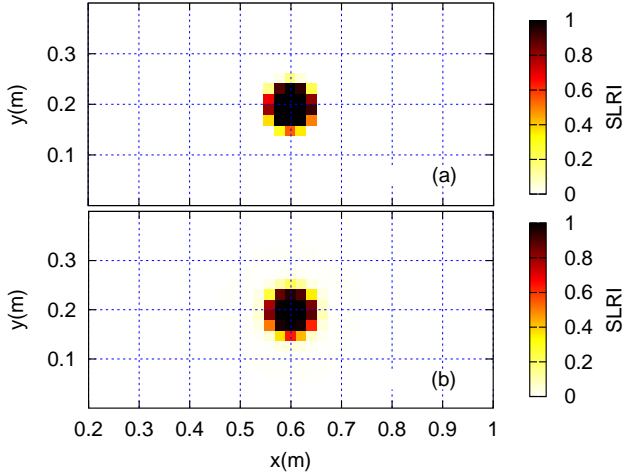}
	\caption{Target (a) and ANN prediction (b) considering the profile of $v_x$ of the fluid generated by an obstacle at the center of the pipe with $\beta = 0.4758$. The color box represents the SLRI. The difference between the target and prediction obstacle is almost imperceptible for the naked eye, the $R^2$ coefficient has reached a value of 0.979.}
	\label{fig:TargetNNObstacleSize20}
\end{figure} 

For measurements made on site B, both $q$ and $v_x$ show $R^2$ values very close to 1.0 for obstacles with diameters greater than 0.05m, that is, for $\beta > 0.25$, see Fig. \ref{fig:TargetNNObstacleSize20} for example, where the target and predicted grid is plotted for an obstacle of $\beta = 0.4758$, achieving a prediction of $R^2=0.979$. However, for small obstacles the accuracy decreases, for example, with $\beta = 0.0488$ we obtained $R^2 = 0.654$ considering $v_x$ and $R^2 = 0.802$ using $q$.

Estimations on the measurement sites A and C for the same diameters, show a similar behavior with small variations on precision, this was expected since the profiles change in time and space. For example, the estimation for the obstacle with a $\beta = 0.0122$, $R^2$ decreases approximately 16\%. This could imply that measuring far away from the obstacle could still produce good estimations of the obstacle's size. Considering the results obtained for this case, it is not possible to establish for which of the physical variables ($v_x$ or $q$) the ANN shows a better performance.

\begin{figure}
	\hspace*{-2cm}
	\includegraphics[width = 11cm]{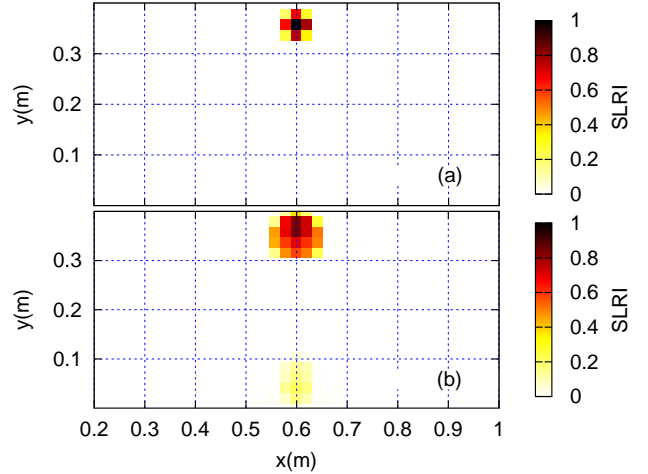}
	\caption{Target (a) and ANN prediction (b) considering the profile of $v_x$ of the fluid generated by an obstacle at $y =  0.35$m with $\beta = 0.122$. Contrasting with Fig. \ref{fig:TargetNNObstacleSize20}, we observe how the ANN got confused and indicates a low SNRI at the bottom of the pipe, resulting in a prediction with $R^2 = 0.09$.}
	\label{fig:YTargetNNObstacleSize11}
\end{figure} 

\begin{figure}
	\includegraphics[width = 9cm]{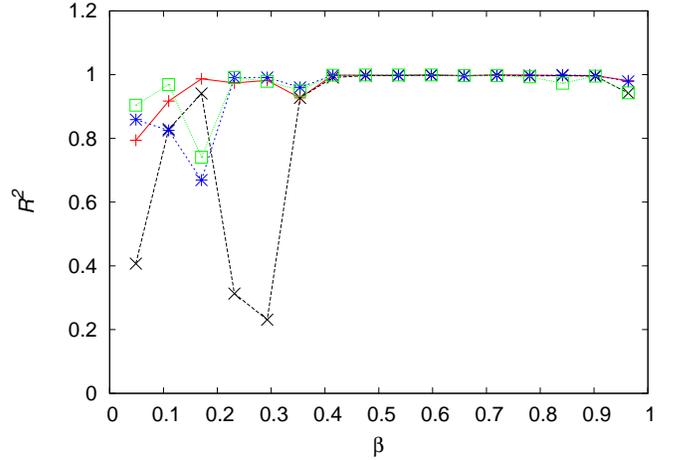}
	\caption{ANNs prediction analysis by $R^2$ for several obstacles' sizes and centered at the pipe. Measuring the time series on a single sensor for $v_x$ (+) and $q$ ($\times$) at $y=0.205$m, and three detectors at $y=$ 0.105m, 0.210m and 0.315m of $v_x$ ($\hexstar$) and $q$ ($\square$) on measurement site B. Notice that by considering the time series of three detectors there are values of $R^2$ close to 1 for diameters greater than $\beta = 0.20$, while with a single one this happens for $\beta > 0.25$.}
	\label{fig:TS_1_3pts_Diams_Center_D_Velx_P}
\end{figure}

The Fig. \ref{fig:TS_1_3pts_Diams_Center_D_Velx_P} shows the results with a single sensor at $y=0.21$m, where we obtained values of $R^2$ above 0.8 considering $v_x$. This accuracy decreases when $q$ is considered as input, obtaining the worst prediction with $R^2 = 0.231$. For the case with three sensors, the results improved considerably maintaining a very similar behavior for both physical variables, obtaining the worst prediction with $R^2 = 0.669$ for $\beta=0.183$. Furthermore in case \ref{p:CaseIC}, the values of $R^2>0.9$ for $\beta > 0.25$.

\begin{figure}
	\includegraphics[width = 9cm]{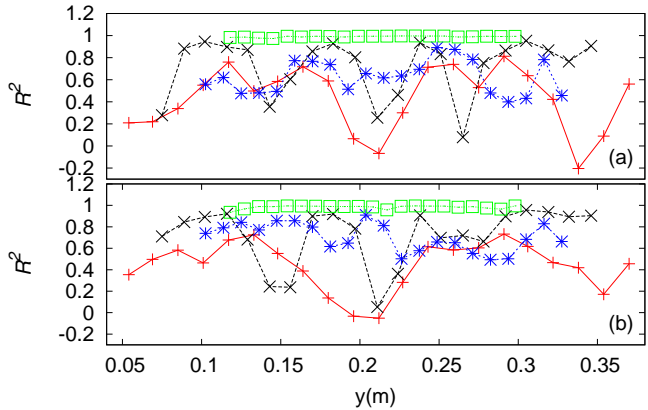}
	\caption{Prediction analysis for obstacles on different positions over the y-axis, by training and testing the ANN with the profiles of $v_x$ (a) and $q$ (b) for detectors at measurement site B. The red (+) curve shows the results for the obstacle $\beta = 0.122$, while the black ($\times$), blue ($\hexstar$) and green ($\square$) curves show those with $\beta = 0.244$, $0.366$ and $0.488$ respectively. Values of $R^2$ are similar for both approaches, where the smallest obstacle has the worst prediction at the center and close to the borders of the pipe.}
	\label{fig:Diams_YPos_D_Velx}			
	
\end{figure}

Following the approach of case 2a, where the ANNs are trained with the profiles of $v_x$ or $q$ measured at site B, with the intention to obtain not only an estimation of the obstruction's size, but also its location over the $y$-axis.
As shown in Fig. \ref{fig:Diams_YPos_D_Velx}, the results have an $R^2$ coefficient for the larger obstacles ($\beta \ge 0.488$) are close to 1.0. However, in relation to the obstacle with $\beta = 0.366$ for which the ANN was not trained at all, the worst results have lowered to a value of $R^2=0.356$ for $v_x$ and $R^2=0.243$ for $q$.  Note that both estimations are made with the obstacle near the walls of the pipeline. Meanwhile, the $R^2$ coefficient for the obstacle with diameter $\beta = 0.244$, has its lowest value when it is located at the center of the pipe, with an $R^2=0.66$ and $R^2=-0.033$ using the profiles of $v_x$ and $q$ respectively. 

\begin{figure}
	\includegraphics[width = 9cm]{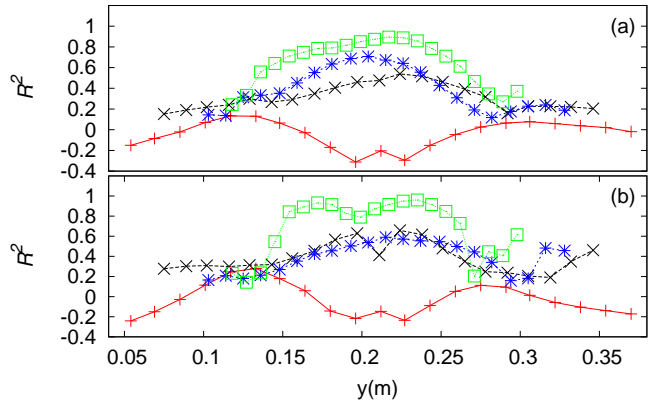}
	\caption{Prediction analysis for obstacles on different positions on the y-axis. The ANNs are trained with the time series of $v_x$ (a) and $q$ (b)at the center of the pipe at measurement site B. The red (+), black ($\times$), blue ($\hexstar$) and green ($\square$) curves show the results for $\beta = 0.122$, $0.244$, $0.366$ and $0.488$ respectively. It is evident the difficulty of the ANN to give a proper prediction about the location of the obstacle with only one sensor, this is more evident for small obstacles close to the borders of the pipe.}
	\label{fig:Diams_YPos_D_TS1pt_Velx}			
\end{figure}

Regarding to case 2b, where the ANN is trained with the time series on a single sensor at the center of the pipe. The ANN is unable to learn the behavior of the flow, mostly for obstacles close to the borders as well as for small ones, as viewed also in case 2a. This is evident from Fig. \ref{fig:Diams_YPos_D_TS1pt_Velx} where we get negative values for the small obstacles $\beta = 0.122$, with a minimum value of $R^2 = -0.312$ when the obstacle is at the center of the pipe. For larger obstacles ($\beta \geq 0.244$) the ANN is more accurate, except when they are near the walls. 

\begin{figure}
	\includegraphics[width = 9cm]{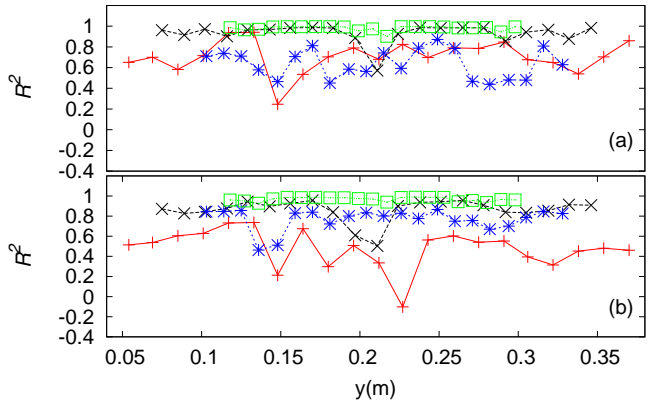}
	\caption{$R^2$ results obtained for obstacles on different positions on the $y$-axis. The ANNs are trained with the time series of $v_x$ (a) and $q$ (b) in three sensors at $y=0.105$m, 0.210m and 0.315m at measurement site B. The red (+), black ($\times$), blue ($\hexstar$) and green ($\square$) curves show the results for obstacles with $\beta = 0.122$, $0.244$, $0.366$ and $0.488$ respectively. Again, the worst adjustment coefficient was obtained with the smallest obstacle, with an $R^2=-0.102$ using $q$ as input data. However, there is remarkable improvement for the other three obstacles sizes compared to the case where a single sensor was used.}
	\label{fig:Diams_YPos_D_TS3pts_Velx}			
\end{figure}

On case 2c, where the time series was generated with three equidistant detectors at $y=0.105$m, 0.210m and 0.315m over measurement site B, the predictions shown in Fig. \ref{fig:Diams_YPos_D_TS3pts_Velx} indicate similar values of $R^2$ for both $q$ and $v_x$. In this scenario, the predictions for the smallest obstacle ($\beta = 0.122$) show an improvement over their counterpart results for case 2b (Fig. \ref{fig:Diams_YPos_D_TS1pt_Velx}). For the obstacle with $\beta = 0.366$ all values of $R^2$ are above 0.4 using $v_x$ or $q$. Meanwhile, for the obstacles with $\beta = 0.244$ and $\beta = 0.488$ the prediction grid shows a great fit respect to the target grid in almost all the locations, with $R^2$ above 0.8. However, the obstacle with $\beta = 0.122$ and located at the center of the pipe has a value of $R^2= -0.102$ using the profile of $q$. 

In general for case \ref{p:CaseII}, the worst adjustment is obtained for the smaller obstacles. This confusion of the ANNs is not surprising, since the pipe boundaries disturb even more the fluid flow generated after the obstacle when it is located near the walls than those scenarios where it is at the center. The boundary layer is affected by the viscous nature of the flow and if the obstacle is very small, the effect it causes to the flow is counteracted by viscous forces, which implies that when we measure far from the obstacle (for example in measurement site B), the flow practically behaves as if there was no obstacle at all. Furthermore, smaller obstacles, regardless of their distance from the pipe boundaries, do not significantly affect the pattern of the flow that is generated behind them, so that the ANN can not characterize them correctly. As it can be seen in case 2b, a single sensor is not enough to obtain a good estimate and also reduces accuracy with respect to case 2a. By increasing to three sensors in case 2c, we obtained a considerable improvement in terms of $R^2$ values.

The results from cases \ref{p:CaseI} and \ref{p:CaseII} show that as the number of measurement points in both space and time increases, the obstacle's size prediction of the ANN improves considerably. For example, comparing cases $1a$ where the ANNs use 83 values in space and only one in time produces, an accuracy equivalent to case $1c$ where three measurements in space and 300 in time are used. A similar conclusion can be achieved by comparing cases 2a and 2c. In other words, the ANN for the case $1a$ performs better than in case $1c$, since \ref{p:CaseIA} requires fewer measurements over time. However, in a practical sense, having a measurement site with only three sensors can be more desirable than an approach of constructing a measurement site with more than 83 sensors.

In reference to case \ref{p:CaseIII}, the $R^2$ coefficients obtained for each scenario are shown in Table \ref{detD}. Here we observe the persistent problem for the smallest obstacle ($\beta = 0.0122$), with an $R^2=-23.753$ when $v_x$ is used, or even worst for $q$ with $R^2=-246.954$. However these particular results have also the second lowest income flow velocity with $v_c = 0.3$m/s. To understand this, compare the results obtained in case \ref{p:CaseIA}, where the associated $v_c$ is always equal to 1.5m/s. That is a decrease in accuracy is associated not only with the difficulty of the ANN to characterize the flow patterns for tiny obstacles, but because the velocity field of the flow around the obstacle is very small also. In other words, in case \ref{p:CaseIII} we found that having both very small obstacles and income flow velocities, the ratio of inertial to viscous forces within the fluid are very low. Notice the improvement of the results for the same value of $\beta = 0.0122$ with $v_c = 1.35$m/s, obtaining $R^2 = -0.645$ for $v_x$ or $R^2 = -0.680$ for $q$. The same happens for the obstacle with $\beta = 0.0976$, improving from $R^2 = 0.250$ to 0.975, and from $R^2 = -4.518$ to 0.975 for $v_x$ and $q$ respectively when the incoming flow velocity is increased from $v_c = 0.15$m/s to $v_c = 0.75$m/s for both variables. This means, that the ANN learns better as we increase the income fluid flow velocity. In general the accuracy of the predictions increases when $v_x$ is used instead of $q$. It is noteworthy, that for most of the remaining predictions the values of $R^2$ are close to 1, with a similar performance compared to the case \ref{p:CaseIA}. Let us remark, that this shows the flexibility of the ANN to work with different initial flow velocities and employing only ten sensors on both measurement sites. 

We have proved that the ANNs achieve a good performance, no matter if we select the training, validation and prediction sets orderly like in cases \ref{p:CaseI} and \ref{p:CaseII} or randomly like in case \ref{p:CaseIII}. The relevance of this last case is that the ANN is trained not only for different diameters, but also for different income flow velocities, implying more complexity with respect to input information, resulting in a clear improvement in the recognition of the shape of the obstacles. Despite that for $\beta = 0.0122$ are obtained bad results, the ANN was able to obtain $R^2>0.96$ for $\beta \geq 0.0976$ when using $v_x$ as input, except in the particular case of $\beta = 0.0976$ with a low income velocity of $v_c = 0.15\rm{m/s}$, which is ten times less than the $v_c$ used in case \ref{p:CaseIA}. A similar result is obtained when $q$ is used, obtaining in general values of $R^2>0.915$. Note that comparing with case \ref{p:CaseI}, the best results were obtained for $\beta > 0.25$ with $R^2>0.9$ for both $v_x$ and $q$.

\begin{table}[]
\centering
\begin{tabular}{|c|c|c|c|}
\hline
$\beta$  & $v_c$ (m/s) & $R^2$ for $v_x$ & $R^2$ for $q$  \\
\hline
\hline
0.0122  & 0.3 & -23.753 & -246.954  \\ \hline
0.0122  & 1.35  & -0.645 & -0.680   \\ \hline
0.0976  & 0.15  & 0.250 & -4.518   \\ \hline
0.0976  & 0.75  & 0.966 & 0.919   \\ \hline
0.0976  & 1.8 & 0.975 & 0.975   \\ \hline
0.1952  & 1.5 & 0.961 & -0.016   \\ \hline
0.2928  & 0.45  & 0.989 & 0.936   \\ \hline
0.2928  & 1.65 & 0.989 & 0.944    \\ \hline
0.3904  & 0.45  & 0.998 & 0.978   \\ \hline
0.3904  & 1.2  & 0.991 & 0.997    \\ \hline
0.4880  & 0.9  & 0.999 & 0.988   \\ \hline
0.4880  & 1.05  & 0.998 & 0.994   \\ \hline
0.4880  & 1.5 & 0.999 & 0.986  \\ \hline
0.5856  & 1.05  & 0.998 & 0.997   \\ \hline
0.5856  & 1.2  & 0.997 & 0.997   \\ \hline
0.7808  & 0.75  & 0.999 & 0.996  \\ \hline
0.7808  & 1.2  & 0.999 & 0.999   \\ \hline
0.8784 & 1.35  & 0.999 & 0.992   \\ \hline
0.9638 & 0.6  & 0.999 & 0.966   \\ \hline
0.9638 & 0.9  & 0.999 & 0.992   \\ \hline

\end{tabular}
\caption{$R^2$ for the test set, considering ten values of the income fluid flow profile of $v_x$ or $q$ before the obstacle and ten values on B. Excluding the three smallest obstacles, the other results are prominent, with values close to 1.}
\label{detD}
\end{table}

\begin{figure*}
	\includegraphics[width = 18cm]{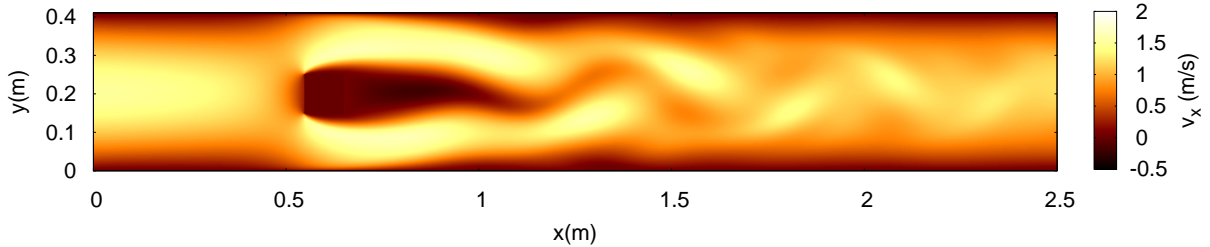}
	\caption{Magnitude of $v_x$, for a simulation of a flow around a square obstacle with an income flow of $v_c =1.5$m/s. The obstacle is located at 0.6m over the $x$-axis. As well as with the cylindrical obstacle, the vortices formed after the square obstacle are driven a long the direction of the flow, generating also a Karman vortex street, but with a different frequency.}
	\label{fig:Final_D21_cuad}
\end{figure*}

\begin{figure}
	\hspace*{-2cm}
	\includegraphics[width =11cm]{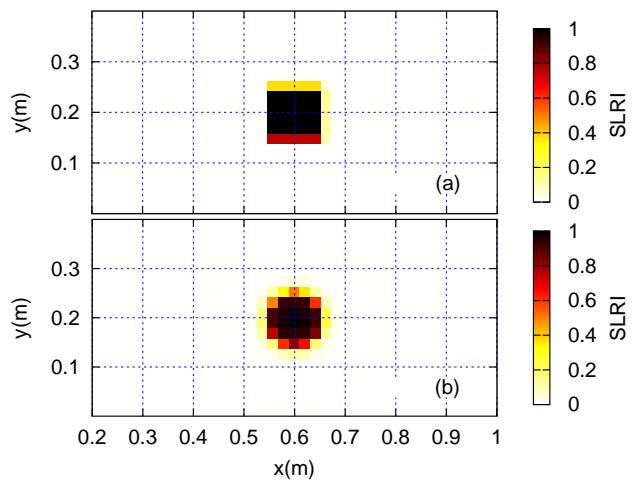}
	\vspace*{-1.3cm}
	\caption{Prediction for an square obstacle at the center of the diameter of the pipe, considering as input of the ANN the time series of $q$ at three different sensors on B. As expected, the ANN considers this obstacle with a shape similar according to those which it was trained for, however it has a comparable size and location of the target.}
	\label{fig:square}
\end{figure}

Hitherto, we have made predictions of the sizes and locations of the obstacles. But we can also test the ability of an already trained ANN to estimate the shape and size of a different obstacle. To show this capacity, we analyze a final experiment, introducing an square obstacle of side 0.1m with the same conditions as those presented in case \ref{p:CaseI}, see Fig. \ref{fig:Final_D21_cuad}. For simplicity, we only present the case for the time series with three different sensors on measurement site B (as in \ref{p:CaseIC}). Despite the predicted shape is like the ones for which the ANN was trained for, we obtained a similar size to the square target, see Fig. \ref{fig:square}, with an outstanding value of $R^2 = 0.93$.

\section{\label{sec:Conclusions} Conclusions}

A series of ANNs have been constructed and trained with the capability to estimate an obstacle's size and location inside a pipe with a specific width of 0.41m, $1/80 \leqslant \beta < 1$ and a range of income flows with characteristic velocities from 0.15m/s to 1.8m/s. The ANNs use as inputs the profile of $v_x$ or $q$ at certain distance of the obstacle. We analyzed several cases varying both the diameter (case \ref{p:CaseI}), the position of the obstacle with respect to the $y$ axis (case \ref{p:CaseII}) and the income fluid velocity (case \ref{p:CaseIII}). 

Based on the specifications used in this work, from the results obtained for case \ref{p:CaseI}, the ANN is highly capable to generate estimations about the obstacle's size when the data supplied is very similar to the one used in the training phase, with results of $R^2>0.9$ for values of $\beta > 0.25$. In case \ref{p:CaseII} something similar was done, considering an obstacle for which the ANN was not trained with any information about the profile or time series of $v_x$ or $q$. Even when there was an evident decrease on accuracy, the ANN was able to estimate not only the shape but also the location of such obstacle with an $R^2>0.6$ in general. And remarkable results for values of $\beta \geq 0.244$ when the obstacles are not too close to the walls of the conduit. Finally in case \ref{p:CaseIII} we have used multiple income flow velocities for each considered blockage, finding out that for low velocities and small obstacles, the ANN predictions have poor accuracy, meanwhile for bigger sizes and greater income flow velocities the accuracy is improved considerably, achieving values of $R^2>0.9$ in general for $\beta \geq 0.0976$.

We found that the ANNs perform better in two different situations. First, when the number of sensors on a measurement site is large as presented in case \ref{p:CaseIA}. Second, when the time series with three sensors is considered as in case \ref{p:CaseIC}. Nevertheless, we think that in a practical sense, is more convenient to extract data on a lapse of time with a few sensors. In the same context, the results by training the ANNs with $v_x$ or $q$ are similar.

Finally, from all cases reviewed the best results are obtained for case \ref{p:CaseIII} given that the ANN can give predictions with different income fluid velocities, meanwhile case \ref{p:CaseI} and \ref{p:CaseII}, were only trained under a single income fluid velocity. Furthermore, case \ref{p:CaseIII} had fewer obstacles in the training set.

Taking into account that in literature one can find works like \cite{Bello}, where they use Modal Analysis to identify a blockage through its location, thickness and depth; or the study done in \cite{Adeleke}, where a transient Pressure-Wave reflection analysis is used to characterize the blockage inside the pipe and for more complex configurations such as the one described as motivation in \cite{Yuan}, we would like to extent our research to develop a numerical tool capable to recognize the shape and depth (its position around the y-axis) of such obstructions. 

\section{Acknowledgments}
This research is partially supported by grant CIC-UMSNH-4.23. We also thank ABACUS Laboratorio de Matem\'aticas Aplicadas y C\'omputo de Alto Rendimiento del CINVESTAV-IPN, grant CONACT-EDOMEX-2011- C01-165873, for providing computer resources.

 
\end{document}